\input harvmac

\def\a{\alpha}
\def\b{\beta}
\def\g{\gamma}
\def\l{\lambda}
\def\d{\delta}
\def\e{\epsilon}
\def\t{\theta}

\def\O{\Omega}
\def\S{\Sigma}

\def\L{\Lambda}

\def\p{\partial}
\def\half{{1\over 2}}
\def\ra{\rightarrow}

\Title{ \vbox{\baselineskip12pt
\hbox{IFT-P.025/2002}}}
{\vbox{\centerline{Massive Superstring Vertex Operator}
\smallskip
\centerline{ in $D=10$ Superspace}}}

\smallskip
\centerline{Nathan Berkovits\foot{e-mail:
nberkovi@ift.unesp.br}}
\smallskip
\centerline{\it 
Instituto de F\'{\i}sica Te\'orica, Universidade
Estadual
Paulista}
\centerline{\it
Rua Pamplona 145, 01405-900, S\~ao Paulo, SP, Brasil}
\bigskip
\centerline{Osvaldo Chand\'{\i}a\foot{e-mail:
ochandia@maxwell.fis.puc.cl}}
\smallskip
\centerline{\it Facultad de F\'{\i}sica, Pontificia
Universidad
Cat\'olica
de Chile}
\centerline{\it Casilla 306, Santiago 22, Chile}

\bigskip

\noindent

Using the pure spinor formalism for the superstring, 
the vertex operator for the first massive states of the open 
superstring is constructed in a manifestly super-Poincar\'e
covariant manner. This vertex operator describes a massive 
spin-two multiplet in terms of ten-dimensional superfields.
 
\Date{April 2002}

\newsec{Introduction}

To study the implications of spacetime supersymmetry for 
the
superstring, it is useful to have a formalism in which 
super-Poincar\'e covariance is manifest.
Although the super-Poincar\'e covariant Green-Schwarz
formalism \ref\GS{M.B.~Green and J.H.~Schwarz,
``Covariant Description of Superstrings,'' Phys. Lett. B136 
(1984)
367.} can be used to classically describe the superstring,
it has not yet been quantized in a covariant manner. 
This prevented the construction of super-Poincar\'e covariant
expressions for massive vertex operators since, unlike
massless vertex operators, massive vertex operators cannot
be obtained from the classical action in a curved background.

Recently,
a new super-Poincar\'e covariant formalism for the 
superstring
has been proposed using a BRST-like operator $Q=\int \l^\a 
d_\a$
where $d_\a$ is the $D=10$ supersymmetric derivative and
$\l^\a$ is a pure spinor variable\ref\superP
{N.~Berkovits, ``Super-Poincar\'e Covariant 
Quantization of the
Superstring,'' JHEP 0004 (2000) 018, hep-th/0001035.}.
In this formalism,
physical vertex operators are defined in a manifestly
super-Poincar\'e covariant manner as states in the 
cohomology
of $Q$. Massless vertex operators have been explicitly
constructed using this formalism and tree amplitudes have
been shown to coincide with Ramond-Neveu-Schwarz amplitudes
\ref\berkval{N.~Berkovits and B.C.~Vallilo,
``Consistency of Super-Poincar\'e Covariant Superstring
Tree Amplitudes,'' JHEP 0007 (2000) 015, hep-th/0004171.}
\ref\relating{
N.~Berkovits, ``Relating the RNS and Pure
Spinor 
Formalisms
for the Superstring,'' JHEP 0108(2001) 026, 
hep-th/0104247.}.

In this paper, the vertex operator for the first massive 
states
of the open superstring will be explicitly constructed in
super-Poincar\'e covariant notation and
shown to describe a massive spin-two multiplet containing
128 bosonic and 128 fermionic degrees of freedom. Although
this construction is guaranteed to succeed because of the
cohomology arguments of \ref\cohomarg
{N.~Berkovits, ``Cohomology in the Pure Spinor Formalism
for the Superstring,'' JHEP 0009 (2000) 046, 
hep-th/0006003\semi
N.~Berkovits and O.~Chand\'{\i}a, ``Lorentz Invariance of 
the Pure
Spinor BRST Cohomology for the Superstring,'' Phys. Lett. 
B514
(2001) 394, hep-th/0105149.}, it is interesting to see how 
the
vertex operator for this massive multiplet is expressed in
terms of ten-dimensional superfields.

\newsec{Physical Vertex Operator}

Physical states in the pure spinor formalism for the open
superstring are defined as ghost-number one states in
the cohomology of $Q=\int \l^\a d_\a$ where 
$\l^\a$ is a pure spinor variable constrained to satisfy
$\l\g^m\l=0$,
\eqn\defdalpha{d_\a=p_\a-\half\g^m_{\a\b}\t^\b\p 
x_m-{1\over
8}\g^m_{\a\b}\g_{m\g\d}\t^\b\t^\g\p\t^\d,}
$\g_m^{\a\b}$ and $\g_{m\a\b}$ are $16\times 16$ symmetric
matrices which are the off-diagonal components of the
$32\times 32$ ten-dimensional gamma matrices,
$[x^m,\t^\a, p_\a]$ for $m=0$ to 9 and $\a=1$ to $16$ are
free worldsheet fields satisfying the OPE's
\eqn\freeope{x^m(y)x^n(z) \ra -\a'\eta^{mn} 
(\log|y-z|+\log|y-\bar z|), 
\quad
p_\a(y)\t^\b(z)\ra {\a'\over{2(y-z)}} \d_\a^\b,}
and $\a'$ is the inverse of the string tension.
One can use
\freeope\ to
show that $d_\a$ is spacetime
supersymmetric and satisfies the OPE's \ref\siege
{W. Siegel, ``Classical Superstring Mechanics,'' Nucl. Phys. B263
(1986) 93.}
\eqn\oped{d_\a(y)d_\b(z)\ra 
-{\a'\over{2(y-z)}}\g^m_{\a\b}\Pi_m(z),\quad
d_\a(y)\Pi^m(z)\ra
{\a'\over{2(y-z)}}\g^m_{\a\b}\p\t^\b(z),}
$$\Pi^m(y)V(z)\ra -{\a'\over{y-z}} \p^m V(z),\quad
d_\a(y)V(z)\ra {\a'\over{2(y-z)}}D_\a V(z),$$
where
$\Pi^m=\p x^m+\half\g^m_{\a\b}\t^\a\p\t^\b$, $V(x,\t)$ is
an arbitrary ten-dimensional superfield, and
$D_\a={\p\over{\p\t^\a}}+\g^m_{\a\b}\t^\b\p_m$
is the supersymmetric covariant derivative which satisfies
$\{D_\a,D_\b\}=2\g^m_{\a\b}\p_m$.

The pure spinor constraint $\l\g^m\l=0$ implies that the
canonical momentum for $\l^\a$, which will be called 
$w_\a$, 
only appears in combinations which are invariant under the
gauge transformation $\d w_\a= (\g^m \l)_\a \L_m$ for 
arbitrary
$\L_m$. This implies that $w_\a$ only appears in the 
Lorentz-covariant
combinations $N_{mn}=\half (w\g_{mn}\l)$ and $J= w_\a 
\l^\a$.
By solving $\l\g^m\l=0$ in terms of unconstrained fields, 
one
can show that $N_{mn}$ and $J$ satisfy the OPE's \relating
\eqn\opeN{ N^{mn}(y)\l^\a(z)\ra {\a'\over{4(y-z)}}
(\g^{mn})^\a{}_\b
\l^\b(z),\quad
J(y)\l^\a(z)\ra {\a'\over{2(y-z)}}\l^\a(z),}
$$N^{kl}(y) N^{mn}(z)\ra -{{3(\a')^2}\over{4(y-z)^2}} 
\eta^{k[n}\eta^{m]l}
+{\a'\over{2(y-z)}}(\eta^{m[l} N^{k]n}(z)-\eta^{m[l} 
N^{k]m}(z)),$$
$$J(y) J(z) \ra -{{(\a')^2}\over{(y-z)^2}}.$$
Furthermore, $\l\g^m\l=0$ implies that 
$N^{mn}$and $J$ satisfy the relation
\eqn\relation{:N^{mn}\l^\a:\g_{m\a\b}-\half
:J\l^\a:\g^n_{\a\b}=\a'\g^n_{\a\b}\p\l^\a(z)}
where the normal-ordered product is defined as
$$:U^A(z)\l^\a(z):
=\oint{dy\over{y-z}} U^A(y) \l^\a(z).$$
To prove \relation, note that $w_\a$ drops out of
the left-hand side because $\l\g^m\l=0$. And the 
coefficient
in the normal ordering
contribution $\a'\g^n_{\a\b}\p\l^\a$ 
can be determined by computing the double
pole of \relation\ with
$J$ using the OPE $J(y)J(z)\ra -{{(\a')^2}\over{(y-z)^2}}$.

When $\a'(mass)^2= 
n$, open superstring vertex operators are constructed
from arbitrary combinations of $[x^m,\t^\a,d_\a, \l^\a, 
N^{mn},J]$
which carry ghost number one and conformal weight $n$ at 
zero
momentum. Note that
$[d_\a,N_{mn},J]$ carry conformal weight one and $\l^\a$ 
carries
ghost number one. For example, the most general 
vertex operator at $(mass)^2=0$
is $V=\l^\a A_\a(x,\t)$ where $A_\a(x,\t)$ is an 
unconstrained
spinor superfield \ref\howe{P.S. Howe, ``Pure Spinor Lines in
Superspace and Ten-Dimensional Supersymmetric Theories,'' Phys. Lett.
B258 (1991) 141.}\superP\ref\grassi{P.A. Grassi, G. Policastro and
P. van Nieuwenhuizen, ``The Massless Spectrum of Covariant Superstrings,''
hep-th/0202123.}.
One can easily check that $QV=0$ and $\d V=Q\Omega$ implies
$\g_{mnpqr}^{\a\b}D_\a A_\b=0$ and $\d A_\a ={\a'\over 2} D_\a \Omega$,
which are the super-Maxwell equations of motion and gauge 
invariances
written in terms of a spinor superfield.

When $\a'(mass)^2=1$, the first massive states of the open superstring 
are
described by the vertex operator
\eqn\VV{V=\p\l^\a A_\a(x,\t)+:\p\t^\b\l^\a B_{\a\b}(x,\t): 
+
:d_\b \l^\a {C^\b}_\a(x,\t):}
$$+:\Pi^m \l^\a H_{m\a}(x,\t):+:J \l^\a E_\a(x,\t):
+
:N^{mn} \l^\a F_{\a mn}(x,\t):$$
where $:U^A \l^\a\Phi_{\a A}(x,\t)(z):$
=$\oint {dy\over{y-z}} U^A(y)~\l^\a(z)
\Phi_{\a A}(z)$ and
$\Phi_{\a A}(x,\t)$ are the various superfields appearing 
in \VV. Note that because of \relation, $V$ is invariant under
the field redefinition
\eqn\vertexgauge{
\d F_{\a mn}=\g_{m\a\b}\L^\b_n-\g_{n\a\b}\L^\b_m,\quad
\d E_\a=-\g^m_{\a\b}\L^\b_m, \quad
\d A_\a=-2\a'\g^m_{\a\b}\L^\b_m.}
As will now be shown, the equations of motion and gauge 
invariances
implied by $QV=0$ and $\d V= Q\Omega$
imply that the superfields $\Phi_{\a A}(x,\t)$ describe a 
massive
spin-two multiplet containing 128 bosonic and 128 fermionic 
degrees
of freedom.

\newsec{Equations of Motion}

Using the OPE's of \oped\ and \opeN, one finds that
\eqn\eqone{{2\over {\a'}} QV=\oint  {{dy}\over{y-z}}[
\p\l^\b(y)\l^\a(z)
(D_\a A_\b(z) +B_{\a\b}(z))-\p\t^\g(y)\l^\a(z)\l^\b(z)D_\b
B_{\a\g}(z)}
$$
-\g_{m\g\b}\l^\g(y)\Pi^m(y)\l^\a(z){C^\b}_\a(z)-d_\g(y)\l^\a(z)\l^\b(z)D_\b
{C^\g}_\a(z)$$
$$
+\g^m_{\g\d}\l^\g(y)\p\t^\d(y)\l^\a(z)H_{m\a}(z)+\Pi^m(y)\l^\a(z)\l^\b(z)D_\b
H_{m\a}(z)$$
$$
-\l^\g(y)d_\g(y)\l^\a(z)E_\a(z)+J(y)\l^\a(z)\l^\b(z)D_\b
E_\a(z)$$
$$
-\half{(\g^{mn})^\g}_\d
\l^\d(y)d_\g(y)\l^\a(z)F_{\a mn}(z)+
N^{mn}(y)\l^\a(z)\l^\b(z)D_\b F_{\a mn}(z)]$$
\eqn\eqtwo{ = -:\p\t^\g \l^\a\l^\b [D_\a
B_{\b\g}-\g^m_{\a\g}H_{m\b}]: + :\Pi^m\l^\a\l^\b[D_\a
H_{m\b}-\g_{m\a\g}{C^\g}_\b]:}
$$-:d_\g\l^\a\l^\b[D_\a{C^\g}_\b+
\d_\a^\g E_\b +\half (\g^{mn})^\g{}_\a F_{\b mn}]:$$
$$+ \l^\a\p\l^\b[D_\a
A_\b+B_{\a\b}+\a'\g^m_{\b\g}\p_m{C^\g}_\a-{\a'\over 2}
D_\b E_\a +{\a'\over 4}
(\g^{mn}D)_\b
F_{\a mn}] $$
$$ +: J \l^\a\l^\b D_\a E_\b: + 
:N^{mn} \l^\a \l^\b D_\a F_{\b mn}:,
$$
where
$:U^A \l^\a\l^\b \S_{\a\b A}(x,\t)(z):$
=$\oint {{dy}\over{y-z}} U^A(y)~\l^\a(z)\l^\b(z)
\S_{\a\b A}(z)$.

Since $\l\g^m\l = 
\l\g^m\p\l$ =0,
$QV=0$ implies that the superfields $\Phi_{\a A}$
satisfy
\eqn\eqs{(\g_{mnpqr})^{\a\b}[D_\a
B_{\b\g}-\g^s_{\a\g}H_{s\b}]=0,}
$$
(\g_{mnpqr})^{\a\b}[D_\a
H_{s\b}-\g_{s\a\g}{C^\g}_\b]=0,$$
$$
(\g_{mnpqr})^{\a\b}[D_\a{C^\g}_\b+ \d^\g_\a E_\b +\half
(\g^{st})^\g{}_\a F_{\b st}]=0,$$
$$
(\g_{mnpqr})^{\a\b}[D_\a
A_\b+B_{\a\b}+\a'\g^s_{\b\g}\p_s{C^\g}_\a-{\a'\over 2}D_\b E_\a +
{\a'\over 4}
(\g^{st} D)_\b F_{\a st} ]$$
$$=
2\a'\g_{mnpqr}^{\a\b} \g^{vwxys}_{\a\b} \eta_{st} K^t_{vwxy},$$
$$
(\g_{mnp})^{\a\b}[D_\a
A_\b+B_{\a\b}+\a'\g^s_{\b\g}\p_s{C^\g}_\a-{\a'\over 2}D_\b E_\a +
{\a'\over 4}
(\g^{st} D)_\b F_{\a st}]$$
$$= 
16\a'\g_{mnp}^{\a\b}
\g^{wxy}_{\a\b} K^s_{wxys},$$
$$\g_{mnpqr}^{\a\b} D_\a E_\b = 
\g_{mnpqr}^{\a\b} (\g^{vwxy}\g_s)_{\a\b} K^s_{vwxy},$$
$$\g_{mnpqr}^{\a\b} D_\a F_{\b}^{st}=
-\g_{mnpqr}^{\a\b} (\g^{vwxy}\g^{[s})_{\a\b} 
K^{t]}_{vwxy},$$
where $K^s_{vwxy}$ is an arbitrary superfield.
The possibility of introducing $K_{vwxy}^t$ into the 
right-hand
side of \eqs\ comes from the 
fact that for arbitrary $K^s_{vwxy}$, 
\eqn\comesf{-:N_{st} \l^\a \l^\b: (\g^{vwxy}\g^{[s})_{\a\b} 
K^{t]}_{vwxy}+:J\l^\a\l^\b:(\g^{vwxy}\g_s)_{\a\b} K^s_{vwxy}}
$$
+ \a'\l^\a \p\l^\b
[2\g^{vwxys}_{\a\b} \eta_{st} K^t_{vwxy}+16\g^{wxy}_{\a\b}K^s_{wxys}]=0,$$
which follows from the identity
\eqn\newid{:N_{st}\l^\a\l^\b:\g^s_{\b\g}-\half
:J\l^\a\l^\b:\g_{t\b\g}={{5\a'}\over 4}\l^\a\p\l^\b \g_{t\b\g}
-{\a'\over 4} \l^\d\p\l^\b (\g_{st})^\a{}_\d  \g^s_{\b\g} .}

To derive \newid, first define 
\eqn\normtwo{:U^A\l^\a\l^\b: = \lim_{w\to z}\oint_{C_w}
 {{dy}\over{y-z}} U^A(y) \l^\a(w)\l^\b(z)
+\lim_{w\to z} \oint_{C_z}
 {{dy}\over{y-z}} U^A(y) \l^\a(w)\l^\b(z)}
where $C_w$ encircles the point $w$ and $C_z$ encircles the point $z$.
Using \relation\ and \opeN, one finds 
\eqn\newfinds{:N_{st}\l^\a\l^\b:\g^s_{\b\g}-\half
:J\l^\a\l^\b:\g_{t\b\g}}
$$=\a'\l^\a\p\l^\b \g_{t\b\g}  +
\lim_{w\to z} \oint_{C_w} 
{{dy}\over{y-z}} (N_{st}(y)\g^s_{\b\g}-\half J(y)\g_{t\b\g}) \l^\a(w)\l^\b(z)$$
$$=\a'\l^\a\p\l^\b \g_{t\b\g}  +
\lim_{w\to z} {1\over{w-z}}[{\a'\over 4}
(\g_{st})^\a{}_\d \l^\d(w) \g^s_{\b\g} -{\a'\over 4}
\l^\a(w)\g_{t\b\g}] \l^\b(z)$$
$$=
\a'\l^\a\p\l^\b \g_{t\b\g}  -
{\a'\over 4} \l^\d \p\l^\b (\g_{st})^\a{}_\d \g^s_{\b\g}
+{\a'\over 4} \l^\a\p\l^\b \g_{t\b\g}.$$

\newsec{Gauge Transformations}

In order to determine the physical content of the equations 
of motion \eqs,
one needs to gauge fix the superfields using the 
gauge transformations implied by $\d V=Q\Omega$, as well as 
the
transformations implied by the field redefinition of 
\vertexgauge. 
Since
the gauge parameter $\Omega$
should have ghost number zero and conformal weight one,  
the most general
gauge parameter is
\eqn\geng{
\O=  :\p\t^\a\O_{1\a}(x,\t): + :d_\a\O^\a_2(x,\t):
+ :\Pi^m\O_{3m}(x,\t):}
$$+ :J\O_4(x,\t): + :N^{mn}\O_{5mn}(x,\t):,$$
where $:U^A \O_A(x,\t):$  
=$\oint {dy\over{y-z}} U^A(y)~ \O_A(z).$

Using the OPE's of \oped\ and \opeN, one finds that
\eqn\resone{{2\over {\a'}} Q\O=\oint 
{{dy}\over{y-z}}
[\p\l^\a(y)\O_{1\a}(z)-\p\t^\b(y)\l^\a(z)D_\a\O_{1\b}(z)
-\g_{m\a\b}\l^\b(y)\Pi^m(y)\O^\a_2(z)}
$$-d_\b(y)\l^\a(z)D_\a\O^\b_2(z)
+\g^m_{\g\d}\l^\g(y)\p\t^\d(y)\O_{3m}(z)+\Pi^m(y)\l^\a(z)D_\a\O_{3m}(z)$$
$$
-\l^\g(y)d_\g(y)\O_4(z)+J(y)\l^\a(z)D_\a\O_4(z)$$
$$
-\half{(\g^{mn})^\g}_\d
\l^\d(y)d_\g(y)\O_{5mn}(z)+N^{mn}(y)\l^\a(z)D_\a\O_{5mn}(z)]$$
\eqn\restwo{=
\p\l^\a[\O_{1\a}+\a'\g^m_{\a\b}\p_m\O^\b_2-{\a'\over 2}
D_\a\O_4-{\a'\over 4}{(\g^{mn})^\b}_\a
D_\b\O_{5mn}]}
$$+:\p\t^\b\l^\a[-D_\a\O_{1\b}+\g^m_{\a\b}\O_{3m}]:$$
$$+
:d_\b\l^\a[-D_\a\O^\b_2-\d^\b_\a\O_4-\half{(\g^{mn})^\b}_\a\O_{5mn}]:$$
$$+:\Pi^m\l^\a[D_\a\O_{3m}-\g_{m\a\b}\O^\b_2]:$$
$$+: J\l^\a D_\a \O_4: + :N^{mn}\l^\a D_\a\O_{5mn}:.$$
So $\d V={2\over {\a'}} Q\O$ implies the following
gauge transformations
for the superfields in \VV :
\eqn\gtransf{\d
A_\a=\O_{1\a}+\a'\g^m_{\a\b}\p_m\O^\b_2-{\a'\over 2}
D_\a\O_4-{\a'\over 4}{(\g^{mn})^\b}_\a
D_\b\O_{5mn},}
$$\d B_{\a\b}=-D_\a\O_{1\b}+\g^m_{\a\b}\O_{3m},$$
$$\d
{C^\b}_\a=-D_\a\O^\b_2-\d^\b_\a\O_4-\half{(\g^{mn})^\b}_\a\O_{5mn},$$
$$\d H_{m\a}=D_\a\O_{3m}-\g_{m\a\b}\O^\b_2,$$
$$\d E_\a=D_\a \O_4,$$
$$\d F_{\a mn}=D_\a\O_{5mn}.$$

\newsec{Massive Spin-Two Multiplet}

In this section, we shall show that the equations of motion 
of \eqs\ and the gauge transformations of \gtransf\ and \vertexgauge\
imply that the 
superfields appearing in \VV\
describe a spin-two multiplet with $(mass)^2={1\over{\a'}}$.
Note that the 128 bosonic 
and
128 fermionic component fields in a massive spin-two 
multiplet consist
of 
a traceless symmetric tensor $g_{mn}$, 
a three-form $b_{mnp}$, and a
spin-3/2 field
$\psi_{m\a}$ satisfying the equations:
\eqn\field{
\eta^{mn} g_{mn}= \p^m g_{mn}=\p^m b_{mnp}=\p^m
\psi_{m\a}=\g^{m\a\b}\psi_{m\b}=0. }
These ten-dimensional component fields can be understood
as Kaluza-Klein modes of an eleven-dimensional supergravity
multiplet.

The 
first equation of motion of \eqs\ implies that
$\l^\a\l^\b\l^\g D_\a B_{\b\g}=0$ where 
$\l^\b\l^\g B_{\b\g}= (\l\g^{mnpqr}\l)B_{mnpqr}$. As
discussed in reference
\ref\particle{N.~Berkovits,
``Covariant Quantization of the Superparticle Using Pure 
Spinors,''
JHEP 0109 (2001) 016, hep-th/0105050.}, this
is the same 
equation of motion as for the super-Maxwell antifield
$A^*_{\b\g}= A^*_{mnpqr} (\g^{mnpqr})_{\b\g}$.
But up to the gauge transformation
$\d A^*_{mnpqr} = \g_{mnpqr}^{\b\g} D_\b \L_\g$, 
$\l^\a\l^\b\l^\g D_\a A^*_{\b\g}=0$
has only massless solutions. Since the fourth equation of 
\eqs\ implies at $\a'=0$ that $\g^{\a\b}_{mnpqr}(B_{\a\b}+D_\a A_\b)=0$, 
$B_{mnpqr}$ has no massless 
solutions. So  
$\l^\a\l^\b\l^\g D_\a B_{\b\g}=0$ implies that
$B_{mnpqr} = \g_{mnpqr}^{\b\g} D_\b \L_\g$
for some $\L_\g$.
Using the gauge parameters $\O_{1\a}$ and $\O_{3m}$, one can
therefore gauge $B_{\a\b}=\g^{mnp}_{\a\b} B_{mnp}$. Note
that this gauge-fixing condition
still leaves gauge invariances parameterized by $\O_{1\a}$
that satisfy $\g_{mnpqr}^{\a\b} D_\a \O_{1\b}=0$.

Plugging 
$B_{\a\b}=\g^{mnp}_{\a\b} B_{mnp}$
into the first equation of \eqs, one
finds that  
\eqn\bone{
(\g^s\g_{mnpqr})_\g{}^\a
H_{s\a}=(-\g^{stu}\g_{mnpqr})_\g{}^\a D_\a B_{stu},}
which implies that 
\eqn\BasH{D_\a B^{mnp}=
\g^{[m}_{\a\b} Z^{np]\b} -{1\over {48}} (\g^{[mn})_\a{}^\b 
H^{p]}_\b
+ \g^{mnp}_{\a\g} Y^\g}
for some $Y^\g$ and $Z^{np\b}$ satisfying
$Z^{np\b}\g_{p\a\b} =0$. 
It will now be argued that \BasH\ implies that $B_{mnp}$
describes a massive spin-two multiplet whose mass will be 
determined
by the fifth equation of \eqs.

To analyze the physical content of \BasH, it will be useful
to choose a reference frame in which the spatial momenta
$k_a=0$ for $a=1$ to 9 where
the indices $[a,b,c,...]$ denote spatial directions.\foot
{Although the analysis would
be more complicated, the physical content of \BasH\
could also be covariantly derived by applying combinations of $D_\a$
and using $\g$-matrix identities.} 
This reference frame is always possible since the fifth equation
of \eqs\ at $\a'=0$ implies that $\g_{mnp}^{\a\b}(B_{\a\b}+D_\a A_\b)=0$,
so $B_{mnp}$ has no massless solutions.
The spatial
polarizations of \BasH\ imply that 
\eqn\lcB{D_\a B^{abc} = \g_{\a\b}^{[a} S^{bc]\b}}
for some superfield $S^{bc\b}$.
To show that $B^{bcd}$ describes a massive spin-two multiplet, recall
that massive representations of $D=10$ supersymmetry (or massless
representations of $D=11$ supersymmetry) are described by the
states $\O^P_A$ where $P$ indices range
over the 128 bosonic and 128 fermionic components of the smallest
$SO(9)$ supersymmetric multiplet and $A$ indices
describe the degeneracy of the ``ground'' state. Note that supersymmetry
transformations act only on the $P$ index and leave the $A$ index 
invariant.

Defining $b^{bcd}= B^{bcd}|_{\t=0}$, the indices $[bcd]$
on $b^{bcd}$ could in principle come from contractions of $P$ indices
with $A$ indices. But the constraint of \lcB\ implies that the
supersymmetry transformation of $b^{bcd}$ is
$\d b^{bcd}= (\e \g^{[b} S^{cd]})|_{\t=0}$, which implies through
the supersymmetry transformation of the $P$ index that all indices
in $b^{bcd}$ come from $P$. So the ``ground state'' $\O^P$ is
non-degenerate and $b^{bcd}$ is the three-form of the smallest
$SO(9)$ supersymmetric multiplet. Furthermore, the supersymmetry
transformation of $b^{bcd}$ implies that
\eqn\deffS{S^{bc\b}= (\g^{[b} \Psi^{c]})^\b} 
where $\g_c^{\b\g} \Psi_\g^c=0$ and
$(\Psi_\g^c)|_{\t=0} = \psi_\g^c$ is the spin $3\over 2$ field.
The remaining 44 bosonic degrees of freedom in the $SO(9)$ multiplet
are described by the $\t=0$ components of the superfield
$G^{bc}= D\g^{(b}\Psi^{c)}$ which satisfies $\eta_{bc} G^{bc}=0$.
Since $G^{mn}$ is a spin-two superfield,
one can interpret $H_{n\alpha}$ as a $D=10$ vector-spinor prepotential as
in \ref\gates{S.J. Gates, Jr. and S. Vashakidze, ``On D = 10, N = 1 
Supersymmetry, Superspace Geometry and Superstring Effects,''
Nucl. Phys. B291 (1987) 172.}. Note that in $D=4$, a similar role
is played by a vector prepotential for a massive spin-two superfield
\ref\leite{N. Berkovits and M.L. Leite, 
``First Massive State of the Superstring in
Superspace,'' Phys. Lett. B415 (1997) 144,
hep-th/9709148\semi N. Berkovits and M.L. Leite,
``Superspace Action for the First
Massive States of the Superstring,''
Phys. Lett. B454 (1999) 38,
hep-th/9812153.} 
\ref\gatestwo{I.L. Buchbinder, S.J. Gates, Jr., J. Phillips and
W.D. Linch, ``New 4D N=1 Superfield Theory: Model of Free
Massive Superspin 3/2 Multiplet,'' hep-th/0201096.}.

To complete the proof that $B^{mnp}$ describes the massive spin-two
multiplet of \field, it will now be shown that 
$B^{0bc}=0$ when $k_a=0$ for $a=1$ to 9. Comparing 
\BasH, \lcB\ and \deffS, one finds that 
\eqn\somec{Y^\g=0,
\quad Z^{bc}= h (\g^{[b}\Psi^{c]})^\g, 
\quad H^b_\b= 96(h -1) \Psi^b_\b}
for some constant $h$. And
$(\g_m Z^{mn})_\a=0$ implies that
$Z^{0b\g}=-7 h (\g^0\Psi^b)^\g$.
After using the gauge parameter $\O_2^\a$ to gauge 
$(\g^m H_m)^\a=0$, one learns from \BasH\ that
\eqn\learnb{D_\a B^{0bc}
= (4-16 h) (\g^0\g^{[b}\Psi^{c]})_\a.}
Using similar arguments as before, one can argue that the only
solution to \learnb\ is $B^{0bc}=0$ and
$h={1\over{4}}$.
To prove this, note that $b^{obc}= B^{0bc}|_{\t=0}$ transforms
under supersymmetry as 
$\d b^{obc}= 
(4- 16 h) (\e\g^0\g^{[b}\psi^{c]}).$
But there are no states in $\O^P_A$ which transform in this manner,
so $b^{0bc}$ must vanish.

So \BasH\ implies that $B^{mnp}$ describes
a massive spin-two multiplet. Furthermore, after using the
gauge parameters $\O_4$, $\O_{5mn}$ and \vertexgauge\ to gauge-fix
\eqn\gfc{C^\a{}_\b = (\g^{mnpq})^\a{}_\b C_{mnpq}\quad {\rm and} \quad
\g^{m\a\b} F_{\b mn}=0,}
the first three equations
of \eqs\ imply that $[H_{m\a}, C_{mnpq}, E_\a, F_{\a mn}]$ are
determined from $B^{mnp}$ by the equations
\eqn\ansatz{H^p_\a = {3\over{7}}(\g_{mn}D)_\a B^{mnp},\quad
C_{mnpq}={1\over {48}} \p_{[m} B_{npq]},\quad E_\a=0,}
$$ F_{\a mn}={7\over {16}} \p_{[m} H_{n]\a} -{1\over {16} }
\p^q(\g_{q[m})^\b{}_\a H_{n]\b},$$
and the trace of the seventh equation of \eqs\ implies that 
\eqn\equK{K^s_{mnpq}= {1\over {1920}}(\g_{mnpqu}^{\a\b} 
D_\a F_\b^{su}
-{1\over{72}} \g_{ru[mnp}^{\a\b} \d_{q]}^s D_\a 
F_\b^{ru}).}

Plugging \ansatz\ and \equK\ into
the fourth equation of \eqs\ implies that 
$\g_{mnpqr}^{\a\b}D_\a A_\b=0$,  so one can gauge-fix
$A_\b=0$ using the remaining gauge transformation parameterized
by $\O_{1\b}$. And plugging \ansatz\
and \equK\ into the fifth equation of \eqs\
implies that 
\eqn\mshell{(\p_m\p^m - {1\over{\a'}}) B_{npq}=0} 
so that 
$(mass)^2={1\over{\a'}}$. Finally,
the sixth equation
and the traceless part of the seventh equation of \eqs\ provide no new 
information,
as can be seen from the fact that if the first five 
equations of \eqs\
are satisfied, 
\eqn\satis{QV= :J(\l\g^{mnpqr}\l): S_{mnpqr} + 
:N_{st}(\l\g^{mnpqr}\l):
T^{st}_{mnpqr}}
for some $S_{mnpqr}$ and traceless $T^{st}_{mnpqr}$.
But $Q^2 =0$ implies that 
\eqn\sattwo{0=Q[ :J(\l\g^{mnqpr}\l): S_{mnpqr}+ 
:N_{st}(\l\g^{mnpqr}\l):
T^{st}_{mnpqr}]}
$$= -{\a'\over 2}\l^\a d_\a 
(\l\g^{mnqpr}\l) S_{mnpqr}+ {\a'\over 4} (\l\g_{st}d) 
(\l\g^{mnpqr}\l)
T^{st}_{mnpqr} + ...$$
where $...$ does not involve $d_\a$. So $Q^2=0$ implies 
that
$S_{mnpqr}=T_{mnpqr}^{st}=0$.

So it has been shown that the vertex operator of 
\VV\ describes a spin-two multiplet with $(mass)^2={1\over{\a'}}$
in terms of ten-dimensional superfields.

\vskip 15pt
{\bf Acknowledgements:} 
We would like to thank 
Brenno Carlini 
Vallilo for his participation in the early stages of this 
project
and the ICTP for their hospitality where part of this work 
was done.
OC would like to
thank FONDECYT grant 3000026 for financial support and 
NB would like to thank
CNPq grant 300256/94-9, Pronex grant
66.2002/1998-9 and FAPESP grant
99/12763-0 for partial financial support.
This research was partially conducted during the period 
that NB
was employed by the Clay Mathematics Institute as a CMI 
Prize Fellow.

\listrefs

\end